# The Edge Electric Field of a Pyroelectric and its Applications

V. Sandomirsky, Y. Schlesinger and R. Levin*

*The Department of Physics Bar-Ilan University, 52900 Ramat-Gan, Israel*

**The College of Judea and Samaria, 44837, Ariel, Israel*

**Abstract**

Following a change of temperature of a pyroelectric (PE), a depolarizing electric field appears both inside the PE, as well as outside its edges, the edge depolarizing electric field (EDEF). The EDEF extends outwards up to a distance of the order of magnitude of the PE width. The mapping and the strength of the EDEF have been calculated and analyzed for the case of a semi-infinite pyroelectric plate. This strong EDEF ($10^4$-$10^5$ V/cm), when penetrating into the surrounding medium, creates a variety of physical effects: inducing electrical current in a semiconductor and affecting its resistance, accelerating charged and neutral particles in vacuum or in a gas, generating electromagnetic waves, modifying optical characteristics by electrooptical and photoelasic effects, generating piezoelectric deformation and more. We show that these EDEF induced effects could serve as a basis for the development of various applications and devices.

Corresponding author: Y. Schlesinger, e-mail: schlesy@mail.biu.ac.il

## 1. Introduction

In a pyroelectric (PE), at constant temperature, the depolarization electric field (DEF) $E_d$, associated with the spontaneous polarization *P,* is screened by charge carriers present in the PE, and/or by the ions of the surrounding ambiance [1,2]. Thus, while *P* is non-zero and its magnitude determined by temperature *T*, the electric field is zero. A change of the PE's temperature, changes *P* and exposes the $E_d \neq 0$, until the screening adjusts to the new state with $E_d = 0$.

The presence of a strong DEF, $10^4 - 10^5$ V/cm, in PE has been demonstrated experimentally by electron and ion emission from ferroelectrics [3, 4, 5]. A recent article [6] reports the observation of ions accelerated up to energies of 200 keV, inducing a D-D reaction in the target, accompanied by generation of neutrons.

However, at present, there are mainly two types of commercial applications of PEs: (1) a temporal change of the PE temperature ($dT/dt \neq 0$) induces a displacement current in the PE, and if its opposite faces (in the direction of the polar axis) are electrically connected to an external conductor, a conductivity current will flow in the external circuit. This is the principle of the pyroelectric IR sensors operation; (2) on the surface of an inhomogeneously heated PE, an electrical potential pattern will form. This is the principle of operation of the vidicon IR thermal imaging [7].

There are currently two generic types of PE sensors, differing by the arrangement of the electrodes relatively to the PE plate [8]. In one type, the electrodes are deposited on opposite sides of PE, perpendicular to the polar axis. In the second type, proposed by A. Handi [8], plane electrodes, with a narrow separating gap, are deposited on the same face of the PE, parallel to the polar axis.

We propose here a different way of a PE application, a configuration in which the PE slab is not a part of the closed current circuit, but, upon a change of temperature, serves solely as the source of the edge depolarizing electric field (EDEF), in its close proximity. This electric field penetrates into a medium, in close contact with the PE, thereby changing its electrical, electro-optical or electro-mechanical characteristics, leading to a whole series of new effects and related potential applications. Thus, depending on a specific application, one would have the possibility to choose an optimal combination of the pyroelectric + substrate.

The schematic model of the layer structure of the PE + (an adjoining medium), is shown in Fig.1.

We calculate here the mapping of the line of force (LF) and the spatial variation of the intensity of EDEF. The results of calculations of EDEF for a simplified model are discussed in Section 2. The details of the calculations are presented in the Appendix. The potential applications are discussed in Section 3.

## 2. The results of the calculation of EDEF

The model system consisting of the PE + adjoining medium (AM) is presented in Fig.1. The PE plate, with a dielectric constant $\varepsilon$, occupies the strip of $-L \leq x \leq L$, $y < 0$, in the lower semi-space. The AM, with a dielectric constant of $\varepsilon_1 << \varepsilon$, occupies the upper semi-space, $y > 0$,. The regions of $x < -L$, and $x > L$, $y < 0$ are filled by a linear dielectric with a dielectric constant $\varepsilon$, equal to that of the PE. This simplification reduces significantly the complexity of the calculations, without affecting the validity or the physical essence of the results.

We will discuss the characteristics of EDEF corresponding to the momentary state following the change of temperature, namely, we consider the spontaneous polarization $\vec{P}(P,0,0)$, and the non-screened DEF and EDEF. The DEF, deep inside the PE, is homogeneous and parallel to the *x*-axis, $E_d = E_x = 4\pi P/\varepsilon$. The EDEF, close to the interface $y = 0$, comprises of both components, $E_x(x,y)$ and $E_y(x,y)$. The general form of $\vec{E}(x,y)$ is similar, but not identical, to the fringe electric field of a parallel-plane capacitor. While in the case of a capacitor the metal plates, at $x = \pm L$, are equipotential surfaces, and the LFs are normal to the plates, in the present case, the boundary planes have a uniform surface charge, $\pm\ P$, and the LFs are not necessarily perpendicular to these planes.

All the following results have been obtained assuming the realistic values $\varepsilon = 1000$, $\varepsilon_l = 10$.

In the analysis we use the dimensionless variables *x,y,z* (normalized to *L*). The solution obviously does not depend on *z*. The electrical field is normalized by *P*, and the potential by $P{\cdot}L$. Since, after this normalization, all the formulas include only $\varepsilon$ and $\varepsilon_l$, the space scale of the field change is *L*.

Our calculations permit to map the lines of force. We discuss here two groups of LFs. One, starts and ends on the inner side of the boundaries of PE at $x = -1$, and $x = +1$, respectively. The second group starts and ends on the outer side, at $x = -1$, and $x = +1$. Three such typical LFs are shown in Fig.2.

The LF1 lies entirely inside the PE slab. The LF starts at the inner boundary at $x = 1$, $y = -0.79$, rising almost vertically alongside the inner side wall up to $y \approx -0.3$. Then, the LF turns left and up, passing the peak point at (0, 0), continuing symmetrically, and ending on the opposite inner wall at the point (-1, - 0.79).

The LF2 starts inside PE at (1, -0.5), rising almost vertically alongside $x = 1$, then, close to the horizontal interface at $y = 0$, turns left, parallel to the axis $y = 0$ and almost coinciding with it. At the point (0.6, 0), it refracts and rises to the peak point at (0, 0.13). Then, the LF continues symmetrically, ending at the inside of the wall at $x = -1$.

The LF3 starts at the outer side of the boundary plane at the point (1, - 0.27), crosses the $y = 0$ plane at the point (1.1, 0). Here it refracts, i.e. its normal component increases according to the ratio of $\varepsilon/\varepsilon_1 = 100$, and the LF enters into the AM up to its peak point at (0, 0.7). Then it continues symmetrically, ending at the outer boundary plane at the point (-1, -0.27).

Figs. 3-5 show the space dependence of the intensity of the tangential component of EDEF, $E_x$.

Fig. 3 shows the $x$-dependence of $E_x$, both inside the PE ($y < 0$) as well as in the AM ($y > 0$), at different depths. All the curves have the same character: they depend weakly on $x$ in the middle part of the PE and increase (in absolute value) at the boundaries. This near boundary variation weakens with increasing depth in the PE.

Fig. 4 shows, on expanded scale, the variation of the tangential field $E_x$ in the AM. Here too, the field is almost uniform in the center region, but decreases on approaching the walls at $x = \pm 1$.

Fig.5 demonstrates that the tangential component inside PE depends only weakly on $y$, decreasing, with characteristic scale of ~ 1, outside of PE. This can be seen also in Fig. 3.

Figs. 6, 7 represent the space dependence of the $y$-component of EDEF. The change of direction, at crossing the plane $x = 0$, is a characteristic of $E_y$. Close to the interface, at $y = 0$, the normal field component, $E_y$, diverges near the boundaries of PE.

A characteristic value of DEF is ~ $P/\varepsilon$. Typical values of EDEF in different regions are listed in the following table:

| | $y < 0$ inside PE | $y < 0$ outside PE | $y > 0$ outside PE |
|---|---|---|---|
| $E_x$ | $\sim 4\pi P/\varepsilon$ | $\sim (P/\varepsilon)\cdot \varepsilon_1/\varepsilon \ll P/\varepsilon$ | $\sim 4\pi P/\varepsilon$ |
| $E_y$ | $\sim (4P/\varepsilon)\cdot \varepsilon_1/\varepsilon \ll P/\varepsilon$ | $\sim (4P/\varepsilon)\cdot \varepsilon_1/\varepsilon \ll P/\varepsilon$ | $P/\varepsilon$ small at $x = 0$ |

If the medium adjacent to PE on its lateral faces (at $|x|>1$ and $y<0$) has a dielectric constant much smaller than that of PE, the lines of force will be "squeezed" strongly toward the PE walls. However, the general structure of the electric field will not change significantly.

## 3. The applications of EDEF

### 3.1. *EDEF in Semiconductors.*

The schematic drawing of such layered structure is shown in Fig. 8a. The semiconductor layer SC (the recipient medium) forms a part of the electrical circuit. It should be emphasized, as stated above, that the sole role of the PE in this structure is to create the EDEF. The PE itself is not a part of the electrical, current-carrying, circuit. We want to follow qualitatively the evolution of the electrical state of the SC after switching-on the EDEF [9].

The EDEF appears, when the PE temperature varies. Assume that the temperature is increasing. There are two routes to increase the temperature, differing markedly in their characteristic time scales. One is the adiabatic "instantaneous" increase of temperature, e.g. by IR laser heating. The other is a "slow" heating by thermal conduction.

The EDEF penetrates into the SC, and redistributes the charge carriers so as to screen EDEF. The characteristic time of this process is of the order of the Maxwell time of the semiconductor, $\tau_m = \varepsilon_1 / 4\pi\sigma$, where $\sigma$ is the semiconductor electroconductivity. The time $\tau_m$ varies in wide limits. When $\sigma$ changes from 1 S to $10^5$ S, the Maxwell time varies between $10^{-10}$ to $10^{-6}$ s. The terms "instantaneous" or "slow" heating are thus defined in comparison with $\tau_m$.

During the initial period of time, shorter than $\tau_m$, EDEF penetrates into the semiconductor to a depth of $\sim L$. The tangential and normal components of EDEF are scaled by $\sim P/\varepsilon$. EDEF induces a current density pulse of $j_s \approx \sigma{\cdot}P/\varepsilon$, flowing through the external circuit closed by the SC. As all the three physical factors ($\sigma$, $P$, $\varepsilon$) are temperature-dependent, this current can vary over wide limits with varying temperature. The current density $j$ depends, via $\sigma$, also on the illumination and on magnetic field. The current density in the semiconductor can exceed the density of pyroelectric current ($j_p = p \cdot dT/dt$, where $p$ is the pyroelectric coefficient, $dT/dt$ is the rate of change of the temperature) by several orders of magnitude.

After the screening is completed, the EDEF resides up to the depth of the order of the Debye screening length, which, typically for a semiconductor, is ~ 1μm. The EDEF will lead to a separation of positive and negative charges of the SC along the $x$-axis, inducing a change of the semiconductor's resistance. This modified resistance state of the semiconductor will be preserved after completing the screening of EDEF by the charge carriers of the SC, resulting in a sort of memory effect.

During the time ~ $\tau_m$, a transient current will flow through the entire volume of the SC. This effect will be larger when the semiconductor layer is thinner than the screening length. To enhance the memory effect, a thin (in comparison with $L$) dielectric buffer layer with a high concentration of traps, can be introduced between the SC and the PE.

To illustrate the above proposition we now develop the theory of a realistic pyroelectric-semiconductor structure device.

The increment of the spontaneous polarization ($\Delta P$), and accordingly the associated EDEF, is given by

$$\Delta P = p \cdot \Delta T \, , \tag{1}$$

where $p$ is the pyroelectric coefficient, and $\Delta T$ is the change of the temperature. Thus, $\Delta P$ is a measure of heating the PE, e.g. by an IR laser, and proportional to the dose of illumination.

Assume that the SC plate is placed into the gap ("sandwich" configuration) between two identical, parallel, plates of PEs (Fig.9). The gap is assumed to be much narrower then the thickness of the PE plates (2*L*). We choose the SC length (*l*) to be less then the PE thickness. Due to this symmetrical arrangement, the EDEF at both lateral faces of the semiconductor (the faces in contact with the PEs) is the same. This EDEF causes a redistribution of the charge carriers in the semiconductor. We will consider the simplest case, when the semiconductor is of *n*-type with totally ionized donor impurities. The positive EDEF will cause an accumulation of the electrons in the left part of the plate and a depletion of electrons in its right part. As a result, the SC plate will become highly resistive. In the following we calculate the dependence of the change of the resistance on $\Delta P$. We label this dependence as the resistance–polarization characteristic (RPC). Since, the polarization is determined by the change of temperature, Eq.(1), RPC can be considered also as a resistance-ΔT characteristic. Evidently, the PE-SC pair acts as a photoresistance, reflecting the change of the PE temperature due to the illumination intensity variations. In contrast from a conventional photoresistor, its resistance <u>increases</u> both with increase of intensity (heating) as well as on decrease of intensity (cooling).

The potential $\varphi$ in the semiconductor plate is described by the Poisson equation

$$\frac{d^2\varphi}{dx^2} = -\frac{4\pi}{\varepsilon_s}\rho; \quad \rho = N - n; \quad n = N \cdot \exp\left(\frac{e\varphi}{kT}\right), \tag{2}$$

where $\rho$ is the space density charge, $\varepsilon_s$ is the dielectric constant of the semiconductor, $N$ is the donor concentration (it is assumed that the donors are fully ionized), $n$ is the free electron concentration, $k$ is the Boltzmann constant. It is assumed that the charge carriers obey the Boltzmann statistics. The potential and the coordinates are referred to the point, where $\rho = 0$. Such point certainly exists, as the semiconductor is neutral on the whole, while its left part is charged negatively and its right part is positive. In dimensionless variables Eq. (2) assumes the form

$$\frac{d^2u}{ds^2} = \exp(u) - 1; \quad u = \frac{\varphi}{kT/e}; \quad s = \frac{x}{L_S}; \; L_S = \sqrt{\frac{\varepsilon_s \cdot kT}{4\pi e^2 N}} \tag{3}$$

The qualitative shape of the potential, *u(s)* is shown in Fig.10. The left end of the semiconductor plate is at $s_1 < 0$, the right one is at $s_2 > 0$. The potentials are respectively, $u(s_1) > 0$, and $u(s_2) < 0$. The boundary conditions applied in the solution of Eq. (3) are

$$\begin{gathered} 4\pi \frac{\Delta P}{\varepsilon} \cdot \frac{\varepsilon_0}{\varepsilon_s} \cdot \frac{e}{(kT/L_S)} = 4\pi \frac{p}{\varepsilon} \cdot \Delta T \cdot \frac{\varepsilon_0}{\varepsilon_s} \cdot \frac{e}{(kT/L_S)} = I = \left.\frac{du}{ds}\right|_{s_1} = \left.\frac{du}{ds}\right|_{s_2}; \\ u(0) = 0; -s_1 + s_2 = S_t = \frac{l}{L_S}, \end{gathered} \tag{4}$$

where $\varepsilon_0$ is the dielectric constant of the gap in PE (e.g., air), $S_t$ is the total dimensionless semiconductor length. Integrating Eq. (3), and applying (4), we obtain Eq.(5) for *u(s)* and the pair of Eqs (6.1) and (6.2) to determine $u_1$ and $u_2$:

$$\int_0^u \left[ I^2 - Q(u_1) + Q(v) \right]^{1/2} dv = -s; \quad Q(u) = u + \exp(u) - 1; \tag{5}$$

$$Q(u_1) = Q(u_2) \tag{6.1}$$

$$\int_{u_2}^{u_1} \left[ I^2 - Q(u_1) + Q(v) \right]^{1/2} dv = S_t \tag{6.2}$$

The system of Eqs. (6.1) and (6.2) is solved numerically. The dependence on $\Delta P$ enters trough *I*. Then, Eq. (5) defines the function *s(u)*, and Eq.(3) gives *n(u)* or *n(s)*.

The relative change of the semiconductor resistance is

$$\theta = \frac{R(\Delta P)}{R(0)} = \frac{N}{l} \int_{x_1}^{x_2} \frac{dx}{n(x)} = \frac{1}{S_t} \int_{u_2}^{u_1} \frac{\exp(-u)du}{\left[ I^2 - Q(u_1) + Q(u) \right]^{1/2}} \tag{7}$$

Thus, the effect is determined by two dimensionless parameters: *I* and $S_t$. The pyroelectric figure of merit (*M*) enters trough *I* as $M = p/\varepsilon$.

Let us present the numerical estimate for a semiconductor sample with following properties: resistivity, $\rho = 10\ \Omega\cdot$cm (assuming an electron concentration of $N = 4.5\cdot10^{14}$ cm$^3$ and mobility 1000 cm$^2$/V·s); $\varepsilon_s = 10$; $T = 300$ K; $l = 5$ mm. Assume also $M = 1$. The calculated graphs are given in Figs. 11 and 12. The minimal $\Delta T = 10$ K corresponds to $\theta = 1.1$, and the maximal $\Delta T = 20$ K corresponds to the depleted bands bend of 0.6 eV.

Fig.11 shows the dependence of the relative change of the resistance on the temperature increment.

Fig.12 depicts the dependencies of the band bend $\Phi_1$ at the left, electron accumulated, "low resistive" end of SC, and of the band bend $\Phi_2$ at the right, depleted, "high resistive" end of SC, on the temperature increment. It can be seen that the potential drop occurs mostly across the high-resistance region s>0, as expected.

We draw the attention to the following points: (1) If the length of the semiconductor is not sufficiently large, then the depletion region can not develop fully; (2) If $\Delta T$ is high enough, then the band bending on the right end of the sample will be also large, creating an inversion layer. Thus, an induced *p-n* junction will be formed, which can be utilized in various ways. The appearance of the inversion layer will not lead to a decrease of the resistance, it will only shift the position of the depleted region inside the SC. However, the dependence of the resistance on $\Delta T$ will saturate; (3) The example represented for $\theta(\Delta T)$ was not optimized. The value of $\theta$ at given $\Delta T$ can be optimized by an appropriate selection of the SC parameters, $T$ and $M$.

**3.2. *The pyroelectric accelerator of charged or neutral polarizable particles.***

In following we demonstrate that EDEF allows to accelerate electrons, ions or neutral polarizable particles and to control their motion in vacuum and in gases.

Consider a hollow cylindrical channel inside the PE, parallel to its polar axis, Fig. 8b. The channel diameter is assumed to be much smaller than the width of the PE, *2L*. Charged particles introduced into the channel will be accelerated by the tangential EDEF, which is of the order of ~ $P/\varepsilon \approx 10^4 - 10^5$V/cm. At a 1 cm length of the channel, the electron will gain an energy of $10^4 - 10^5$ eV within a time ~ $10^{-10}$ s. This time is much less, than any relaxation time of the EDEF due to screening. Similar conclusion is valid also for heavy ions.

Also, using different arrangement of the PE-sections one can control the trajectories of charged particles. Due to the large variety of possible PE-section configurations, there exist a correspondingly large number of potential applications of the EDEF.

The accelerated charged particles will emit electromagnetic radiation due to acceleration, or as a result of hitting a target. In this case, as well, one can consider a diversity of applications. For example, using a channel with periodically corrugated surface of the PE, or periodically alternating polarization, a modulated electromagnetic radiation can be obtained.

Inhomogeneous EDEF will accelerate also neutral polarizable particles or clusters. Different configurations of specifically inhomogeneous EDEF can be formed, for example a tapered diameter channel. More complex space-time configurations of the accelerating electrical field can be realized as well.

**3.3.** ***Piezoelectric*****,** ***Electrooptical and Photoelasic effects***

As mentioned above, the EDEF has a value of ~ $P/\varepsilon \approx 10^4 \div 10^6$ V/cm. This is the typical electrical field intensity for inducing piezoelectric and electrooptical phenomena [11, 12].

In a piezoelectric medium, the EDEF due to the inverse piezoelectric effect, will induce a deformation (Fig.8d) [11], thus transforming heat (e.g. IR light) into a mechanical response. Applying different configurations of a PE or a system of several PEs, one can realize different deformations of the piezoelectric. EDEF allows tuning acoustoelectric units, e.g. the acoustoelectric delay lines [13].

All the well known electrooptical effects, such as the universal quadratic Kerr effect, the linear Pockel's effect, the Stark effect, and the effects in liquid crystals and semiconductors, can be induced by EDEF.

One can also use EDEF for electrooptical modulation. However, its response time will be limited by the relatively slow cooling process. For some of the contemporary pyroelectrics with a pyroelectric coefficient $p$ as high as ~ 1μC/cm$^2$ and $\varepsilon$ ~ 50, according to Eq.(1), even a $\Delta T$ as small as ~ 0.1 K will create an EDEF of an intensity > $10^4$ V/cm.

Using the inverse piezoelectric effect, EDEF induced photoelastic effects should also appear [12].

**3.5. Other potential applications.**

In the following we list some additional potential applications based on the EDEF principle.

**a)** The EDEF of a needle-shaped PE probe can be used to control nanostructures, and to investigate the characteristics of nano-size systems. Based on the fact that ferroelectricity was found to sustain up to a sample size of about 40 Å [14], it is reasonable to presume that pyroelectricity will sustain as well at such dimension. Assuming that it is possible to create a PE tip with a diameter of ~ 100 Å, the EDEF of such PE tip can be used to manipulate nano-particles in space, to change the energy spectrum of nanostructures, thus affecting their electronic properties, or to measure their local characteristics, as e.g., hysteresis loops.

b) Due to energy considerations, a polarizable liquid will be sucked into the channel in PE due to EDEF. One can imagine a variety of devices based of this phenomenon.

c) Consider the electro-caloric effect (ECE) [1, 2], i. e., the change of temperature of a dielectric, due to an applied electric field. The condition for the ECE existence is a temperature dependent dielectric constant. The ECE is quadratic in the electric field, the sign of the ECE being determined by the sign of $\partial\varepsilon/\partial T$. Let $\partial\varepsilon/\partial T < 0$, so that ECE leads to cooling of the dielectric in an electric field. Then, an interesting situation arises, when this dielectric is exposed to the EDEF of a PE: when the temperature of the PE rises, EDEF will appear, and the adjacent dielectric will cool.

## 4. Conclusion

We have proposed here a number of possible applications of the edge electric field of a pyroelectric. In distinction with the standard utilization of a PE, in the present case the sole role of the PE is the creation of the EDEF. This field penetrates into an adjacent medium, inducing in it a variety of effects. We present here the calculated mapping of the EDEF lines-of force as well as the calculated intensity of it's components. We also suggest a variety of devices that can be designed based on these effects.

Nowadays, large efforts are spent to develop PEs with as large as possible pyroelectric coefficient. However, according to the idea presented here, a different approach could be the choice of an optimal "PE – recipient medium" pair.

**Appendix.**

The scheme describing the solution of the problem is shown in Fig.13.

The problem is solved by the method of images [2]. The resulting potential is equivalent to the potential due to two charged semi-infinite planes (the projections of the planes on the negative *x-y* plane are sketched in heavy black lines) with surface charge $\pm P$. Respectively, the image (drawn in grey) semi-infinite planes, at $x = \pm 1, y > 0$ possess a surface charge of $\pm P\times(\varepsilon - \varepsilon_l)/(\varepsilon + \varepsilon_l)$. These charges and the black real charged planes create the potential in the lower semi-space. The auxiliary (light grey) semi-planes ($x = \pm 1, y < 0$) have a surface charge of $\pm P\times 2\varepsilon_l/(\varepsilon + \varepsilon_l)$, and give rise to the potential in the upper semi-space. Thus, there are altogether six charged planes: two real and four auxiliary planes. (It should be noticed that if the regions, $x < -1, y < 0$, and $x > 1, y < 0$ are occupied by a third medium with a dielectric constant $\varepsilon_2$, then 18 planes are required to reach the solution). The solution is obtained by direct calculation of the contributions of the corresponding surface integrals from the charged planes.

The potential Φ(x,y), measured in units of $P\cdot L$, is given by the following formulae:

| | | |
|---|---|---|
| $\underline{y \le 0, x \le -1;}$ | $\Phi(x,y) = \frac{4}{\varepsilon}\cdot\frac{\varepsilon_1}{\varepsilon+\varepsilon_1}\{\frac{1}{2}y\ln\frac{(x-1)^2+y^2}{(x+1)^2+y^2} - (x+1)\arctan\left(\frac{y}{x+1}\right) + +(x-1)\arctan\left(\frac{y}{x-1}\right)\} - \frac{4\pi}{\varepsilon+\varepsilon_1};$ | (A1) |
| $\underline{y \le 0, -1 \le x \le 1;}$ | $\Phi(x,y) = \frac{4}{\varepsilon}\frac{\varepsilon_1}{\varepsilon+\varepsilon_1}\{\frac{1}{2}y\ln\frac{(x-1)^2+y^2}{(x+1)^2+y^2} - (x+1)\arctan\left(\frac{y}{x+1}\right) + +(x-1)\arctan\left(\frac{y}{x-1}\right)\} + \frac{4}{\varepsilon+\varepsilon_1}\pi x;$ | (A2) |
| $\underline{y \le 0, x \ge 1;}$ | $\Phi(x,y) = \frac{4\varepsilon_1}{\varepsilon(\varepsilon+\varepsilon_1)}\{\frac{1}{2}y\ln\frac{(x-1)^2+y^2}{(x+1)^2+y^2} - (x+1)\arctan\left(\frac{y}{x+1}\right) + +(x-1)\arctan\left(\frac{y}{x-1}\right)\} - \frac{4}{\varepsilon+\varepsilon_1}\pi;$ | (A3) |
| $y \ge 0, x \le -1;$ | $\Phi(x,y) = \frac{4}{\varepsilon + \varepsilon_1}\{\frac{1}{2} y \ln \frac{(x-1)^2 + y^2}{(x+1)^2 + y^2} - (x+1)\arctan\left(\frac{y}{x+1}\right) + (x-1)\arctan\left(\frac{y}{x-1}\right)\} - \pi;$ | (A4) |
| $y \ge 0, -1 \le x \le 1;$ | $\Phi(x,y) = \frac{4}{\varepsilon + \varepsilon_1}\{\frac{1}{2} y \ln \frac{(x-1)^2 + y^2}{(x+1)^2 + y^2} - (x+1)\arctan\left(\frac{y}{x+1}\right) + (x-1)\arctan\left(\frac{y}{x-1}\right)\} + \pi x;$ | (A5) |
| $y \ge 0, x \ge 1;$ | $\Phi(x,y) = \frac{4}{\varepsilon + \varepsilon_1}\{\frac{1}{2} y \ln \frac{(x-1)^2 + y^2}{(x+1)^2 + y^2} - (x+1)\arctan\left(\frac{y}{x+1}\right) + (x-1)\arctan\left(\frac{y}{x-1}\right)\} + \pi;$ | (A6) |

*W(x,y)* is the function conjugate to the potential *Φ(x,y)*, and is used to construct the lines of force of the electrical field. The function *W(x,y)* is derived under Cauchy-Riemann conditions :

| | | |
|---|---|---|
| $y \le 0, x \le -1;$ | $W(x,y) = \frac{4\varepsilon_1}{\varepsilon + \varepsilon_1}\{y \cdot \arctan\left(\frac{y}{x-1}\right) - \frac{x-1}{2}\ln\left[(x-1)^2 + y^2\right] - y \cdot \arctan\left(\frac{y}{x+1}\right) + \frac{x+1}{2}\ln\left[(x+1)^2 + y^2\right]\} + const;$ | (A7) |
| $y \le 0, -1 \le x \le 1;$ | $W(x,y) = \frac{4\varepsilon_1}{\varepsilon + \varepsilon_1}\{y \cdot \arctan\left(\frac{y}{x-1}\right) - \frac{x-1}{2}\ln\left[(x-1)^2 + y^2\right] - y \cdot \arctan\left(\frac{y}{x+1}\right) + \frac{x+1}{2}\ln\left[(x+1)^2 + y^2\right] + \frac{\varepsilon}{\varepsilon_1}\pi y\} + const;$ | (A8) |
| $y \le 0, x \ge 1;$ | $W(x,y) = \frac{4\varepsilon_1}{\varepsilon+\varepsilon_1}\{y \cdot \arctan\left(\frac{y}{x-1}\right) - \frac{x-1}{2}\ln\left[(x-1)^2+y^2\right] -$ <br> $-y \cdot \arctan\left(\frac{y}{x+1}\right) + \frac{x+1}{2}\ln\left[(x+1)^2+y^2\right]\} + const;$ | (A9) |
| $y \ge 0, x \le -1;$ | $W(x,y) = \frac{4}{\varepsilon+\varepsilon_1}\{y \cdot \arctan\left(\frac{y}{x-1}\right) - \frac{x-1}{2}\ln\left[(x-1)^2+y^2\right] -$ <br> $-y \cdot \arctan\left(\frac{y}{x+1}\right) + \frac{x+1}{2}\ln\left[(x+1)^2+y^2\right]\} + const;$ | (A10) |
| $y \ge 0, -1 \le x \le 1;$ | $W(x,y) = \frac{4}{\varepsilon+\varepsilon_1}\{y \cdot \arctan\left(\frac{y}{x-1}\right) - \frac{x-1}{2}\ln\left[(x-1)^2+y^2\right] -$ <br> $-y \cdot \arctan\left(\frac{y}{x+1}\right) + \frac{x+1}{2}\ln\left[(x+1)^2+y^2\right] + \pi y\} + const;$ | (A11) |
| $y \ge 0, x \ge 1;$ | $W(x,y) = \frac{4}{\varepsilon+\varepsilon_1}\{y \cdot \arctan\left(\frac{y}{x-1}\right) - \frac{x-1}{2}\ln\left[(x-1)^2+y^2\right] -$ <br> $-y \cdot \arctan\left(\frac{y}{x+1}\right) + \frac{x+1}{2}\ln\left[(x+1)^2+y^2\right]\} + const;$ | (A12) |

The electric field components, $E_x$ and $E_y$, in units of $P$, are:

| | | |
|---|---|---|
| $y \le 0, x \le -1;$ | $E_x = -\frac{4}{\varepsilon}\frac{\varepsilon_1}{\varepsilon+\varepsilon_1}\left[\arctan\left(\frac{y}{x-1}\right) - \arctan\left(\frac{y}{x+1}\right)\right];$ <br> $E_y = -\frac{4\varepsilon_1}{\varepsilon(\varepsilon+\varepsilon_1)}\ln\frac{(x-1)^2+y^2}{(x+1)^2+y^2};$ | (A13) |
| $y \le 0, -1 \le x \le 1;$ | $E_x = -\frac{4}{\varepsilon}\frac{\varepsilon_1}{\varepsilon+\varepsilon_1}\left[\arctan\left(\frac{y}{x-1}\right) - \arctan\left(\frac{y}{x+1}\right)\right] - \frac{4\pi}{\varepsilon+\varepsilon_1};$ <br> $E_y = -\frac{2}{\varepsilon}\frac{\varepsilon_1}{\varepsilon+\varepsilon_1}\ln\frac{(x-1)^2+y^2}{(x+1)^2+y^2};$ | (A14) |
| $y \le 0, x \ge 1;$ | $E_x = -\frac{4}{\varepsilon}\frac{\varepsilon_1}{\varepsilon+\varepsilon_1}\left[\arctan\left(\frac{y}{x-1}\right) - \arctan\left(\frac{y}{x+1}\right)\right];$ | (A15) |
| | $E_y = -\frac{2}{\varepsilon}\frac{\varepsilon_1}{\varepsilon+\varepsilon_1}\ln\frac{(x-1)^2+y^2}{(x+1)^2+y^2}$; | |
| $y \ge 0, x \le -1$; | $E_x = -\frac{4}{\varepsilon+\varepsilon_1}\left[\arctan\left(\frac{\mathrm{y}}{\mathrm{x-1}}\right)\text{-}\arctan\left(\frac{\mathrm{y}}{\mathrm{x+1}}\right)\right]$; <br> $E_y = -\frac{2}{\varepsilon+\varepsilon_1}\ln\frac{(x-1)^2+y^2}{(x+1)^2+y^2}$; | (A16) |
| $y \ge 0, -1 \le x \le 1$; | $E_x = -\frac{4}{\varepsilon+\varepsilon_1}\left[\arctan\left(\frac{\mathrm{y}}{\mathrm{x-1}}\right)\text{-}\arctan\left(\frac{\mathrm{y}}{\mathrm{x+1}}\right)\right] - \frac{4\pi}{\varepsilon+\varepsilon_1}$; <br> $E_y = -\frac{2}{\varepsilon+\varepsilon_1}\ln\frac{(x-1)^2+y^2}{(x+1)^2+y^2}$; | (A17) |
| $y \ge 0, x \ge 1$; | $E_x = -\frac{4}{\varepsilon+\varepsilon_1}\left[\arctan\left(\frac{\mathrm{y}}{\mathrm{x-1}}\right)\text{-}\arctan\left(\frac{\mathrm{y}}{\mathrm{x+1}}\right)\right]$; <br> $E_y = -\frac{2}{\varepsilon+\varepsilon_1}\ln\frac{(x-1)^2+y^2}{(x+1)^2+y^2}$. | (A18) |

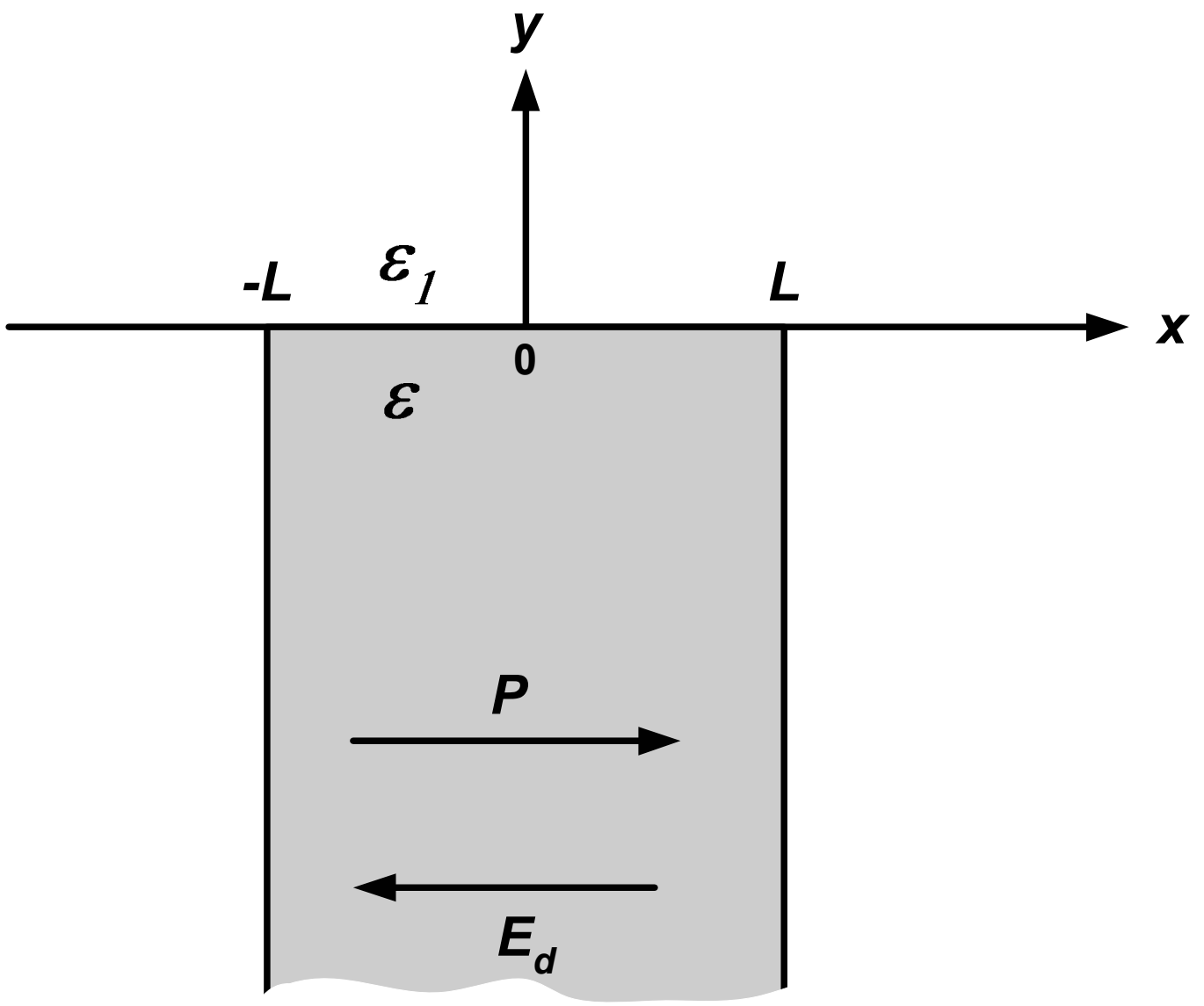

Figure 1. Schematic model of the structure PE+adjoining medium (AM). The PE occupies the region $-L \leq x \leq +L$, $y < 0$. AM occupies the region $y > 0$.

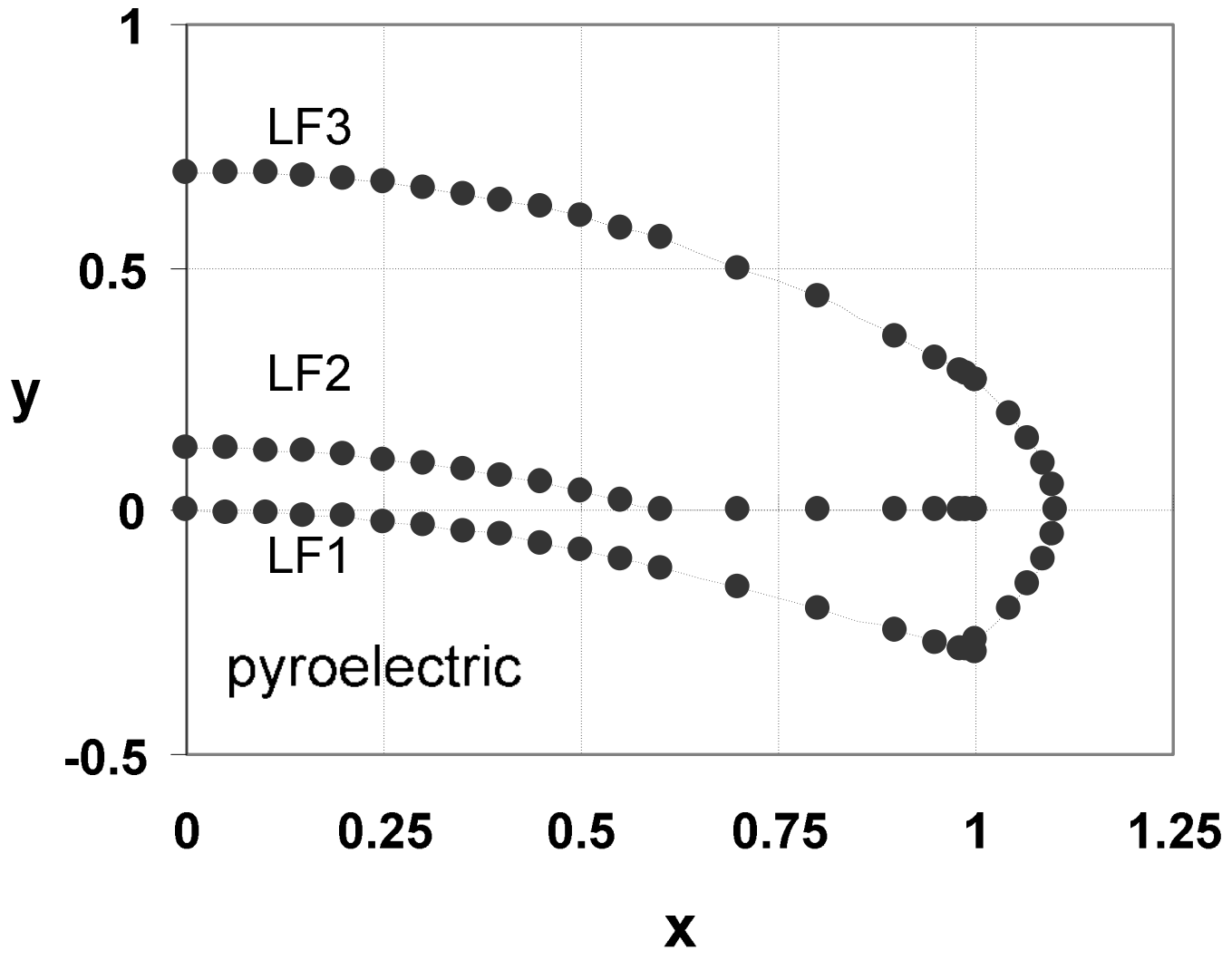

Figure 2. Mapping of three typical lines-of-force. LF1 lies entirely inside the PE slab, starting at the inner boundary at point (1,- 0.79). LF2 starts inside PE at (1, -0.5). LF3 starts at the outer side of the boundary plane at the point (1, - 0.27),

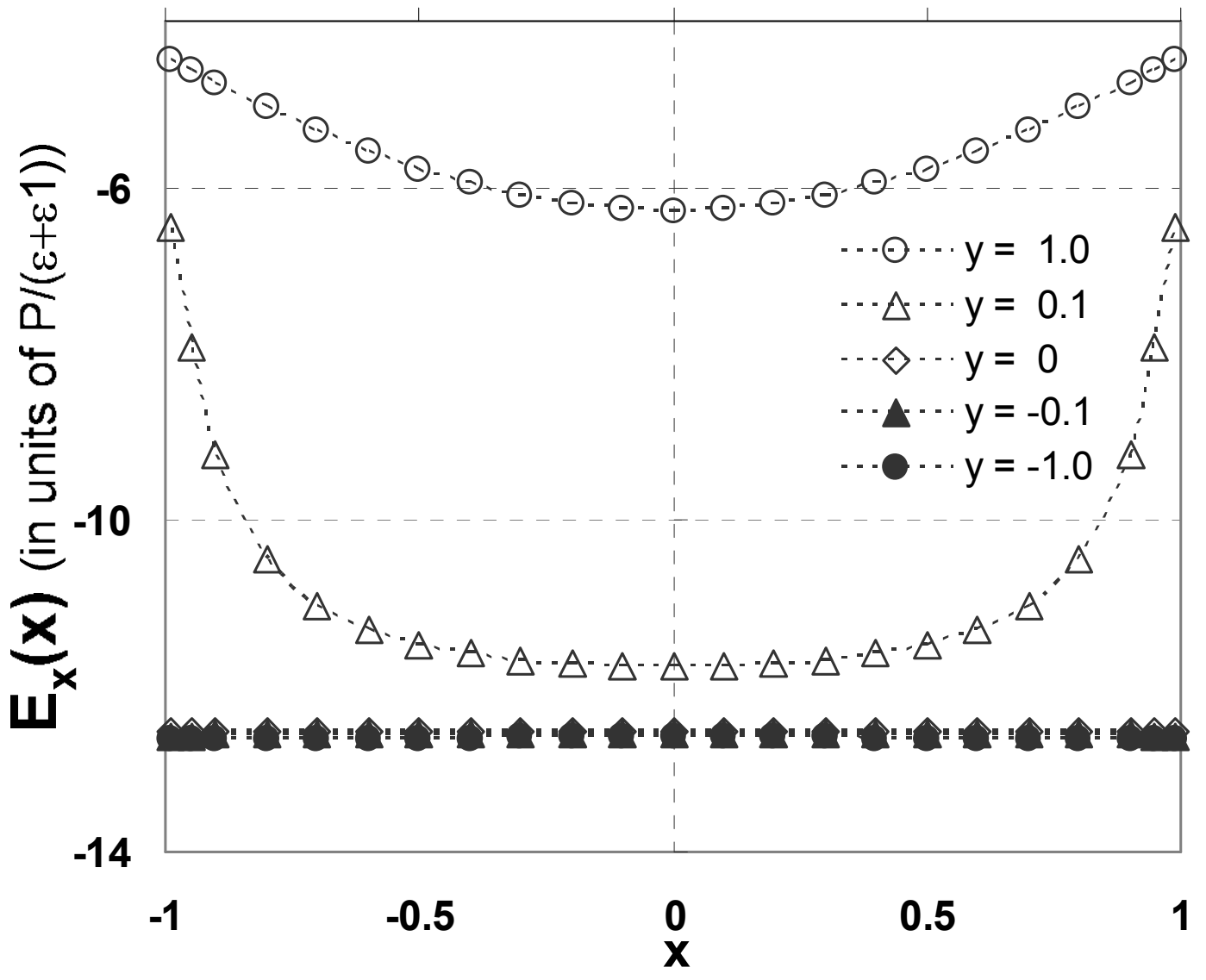

Figure 3. The $x$-dependence of the tangential component $E_x$ of EDEF at different depths in the PE ($y < 0$) and in the AM ($y > 0$).

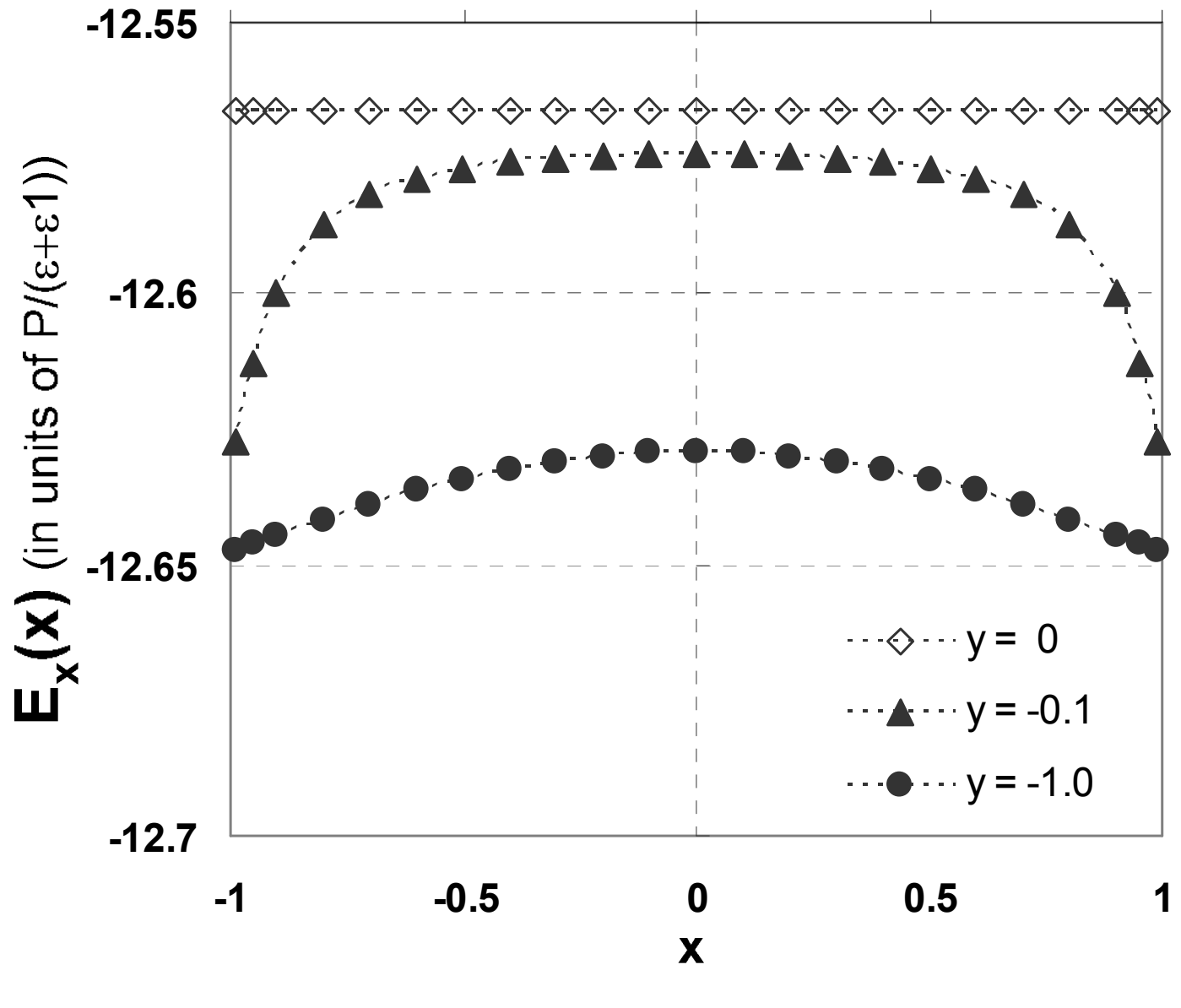

Figure 4. Expanded scale view of the $x$-dependence of the tangential component $E_x$ of EDEF at different depths in the PE ($y < 0$).

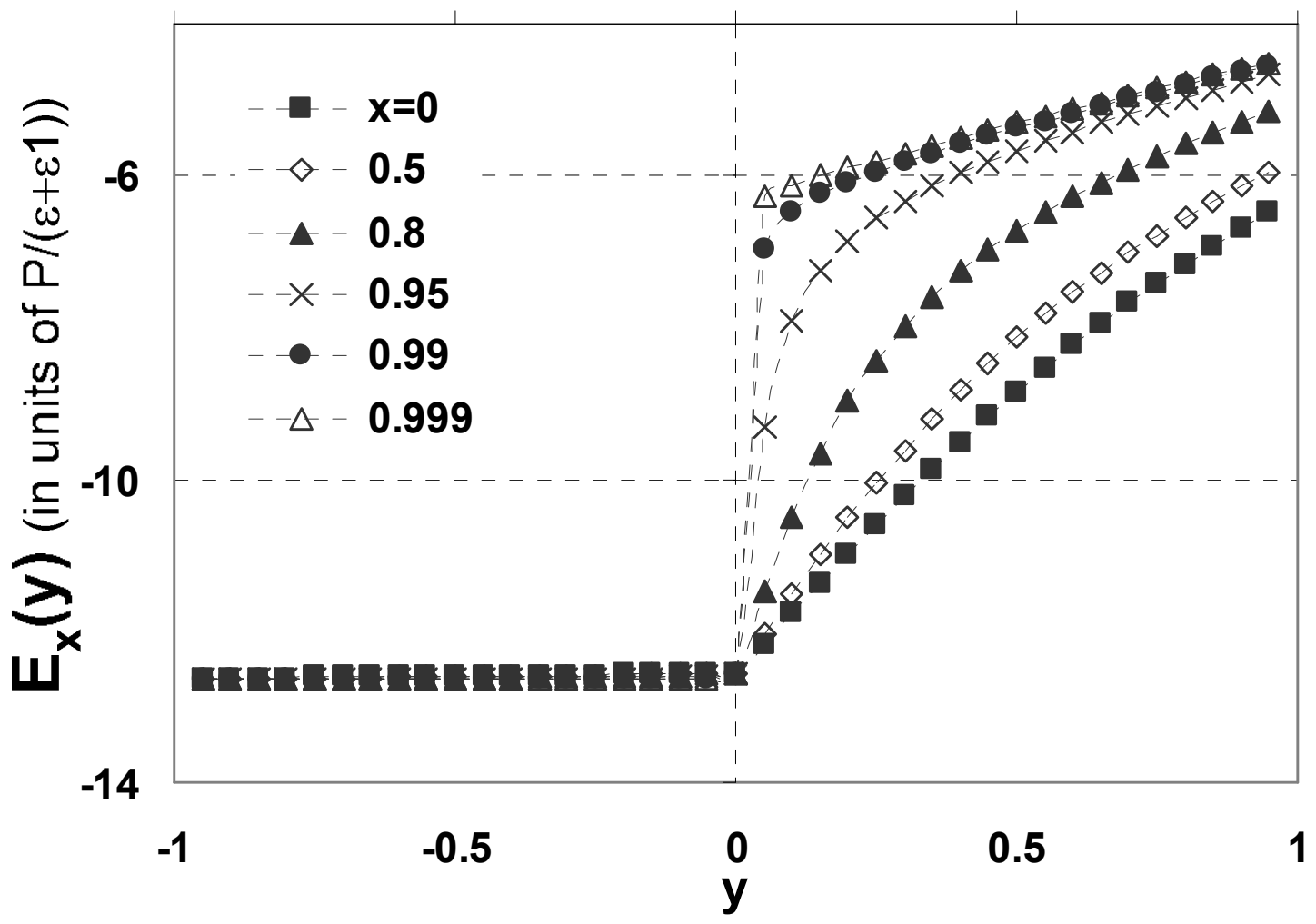

Figure 5. The $y$-dependence of the tangential component $E_x$ of EDEF at several chosen values of $x$.

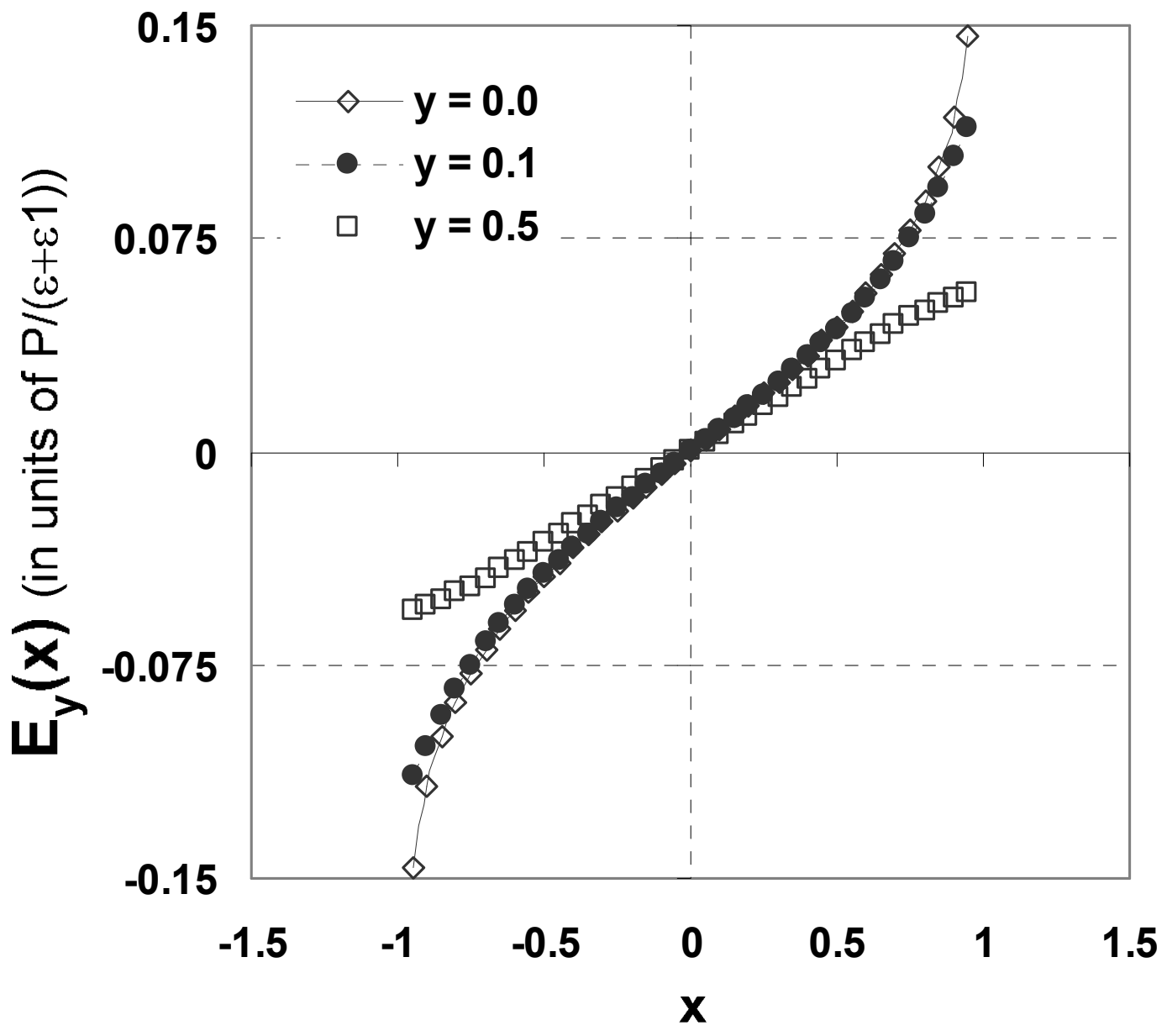

Figure 6. The $x$-dependence of the normal component $E_y$ of EDEF at different depths in the AM ($y > 0$).

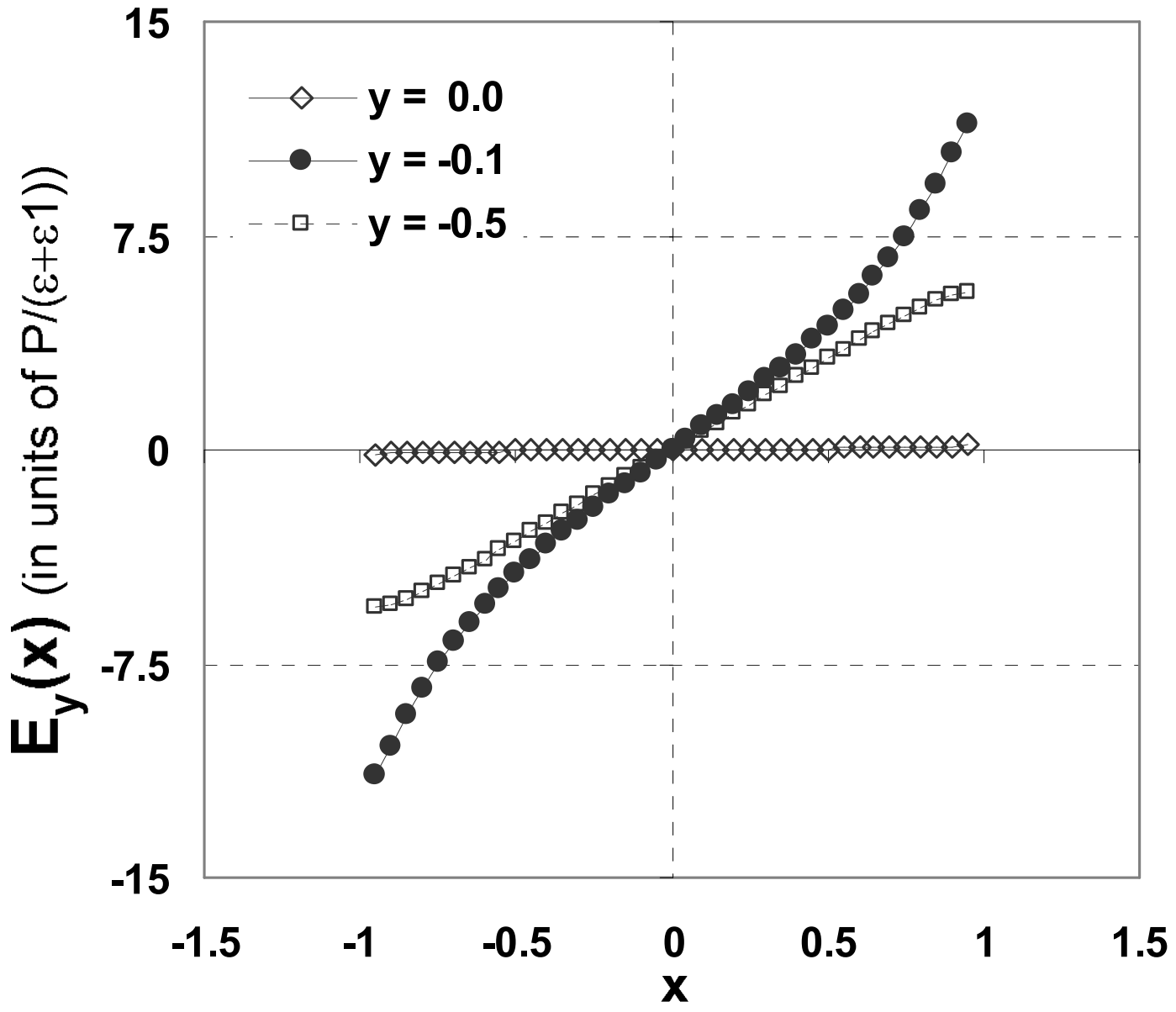

Figure 7. The $x$-dependence of the normal component $E_y$ of EDEF at different depths in the PE ($y < 0$).

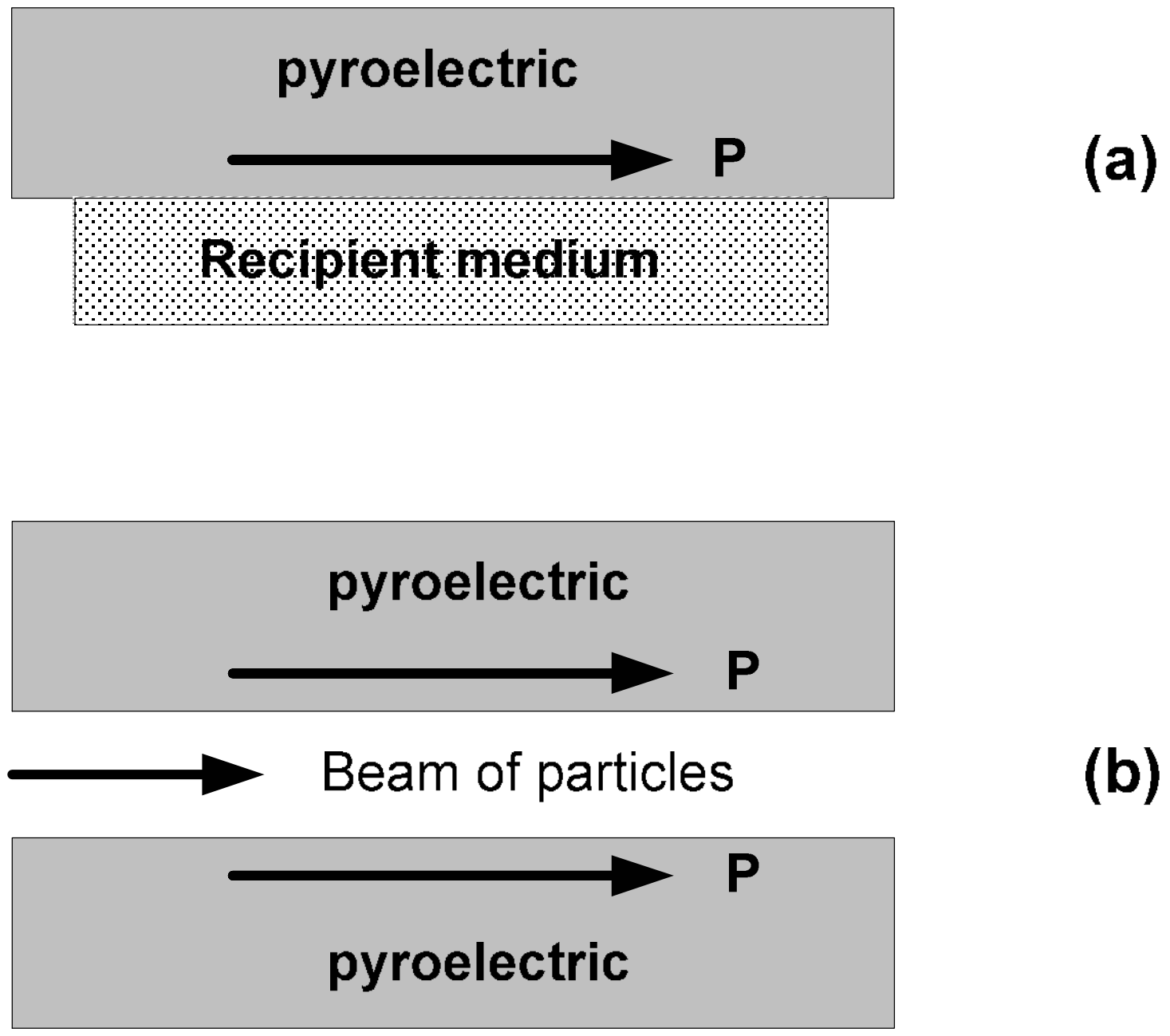

Figure 8. (a) - Schematics of the layered EDEF structure PE + recipient medium (semiconductor, electro-optic medium, piezoelectric medium etc.).
(b) – schematics of the EDEF particle accelerator.

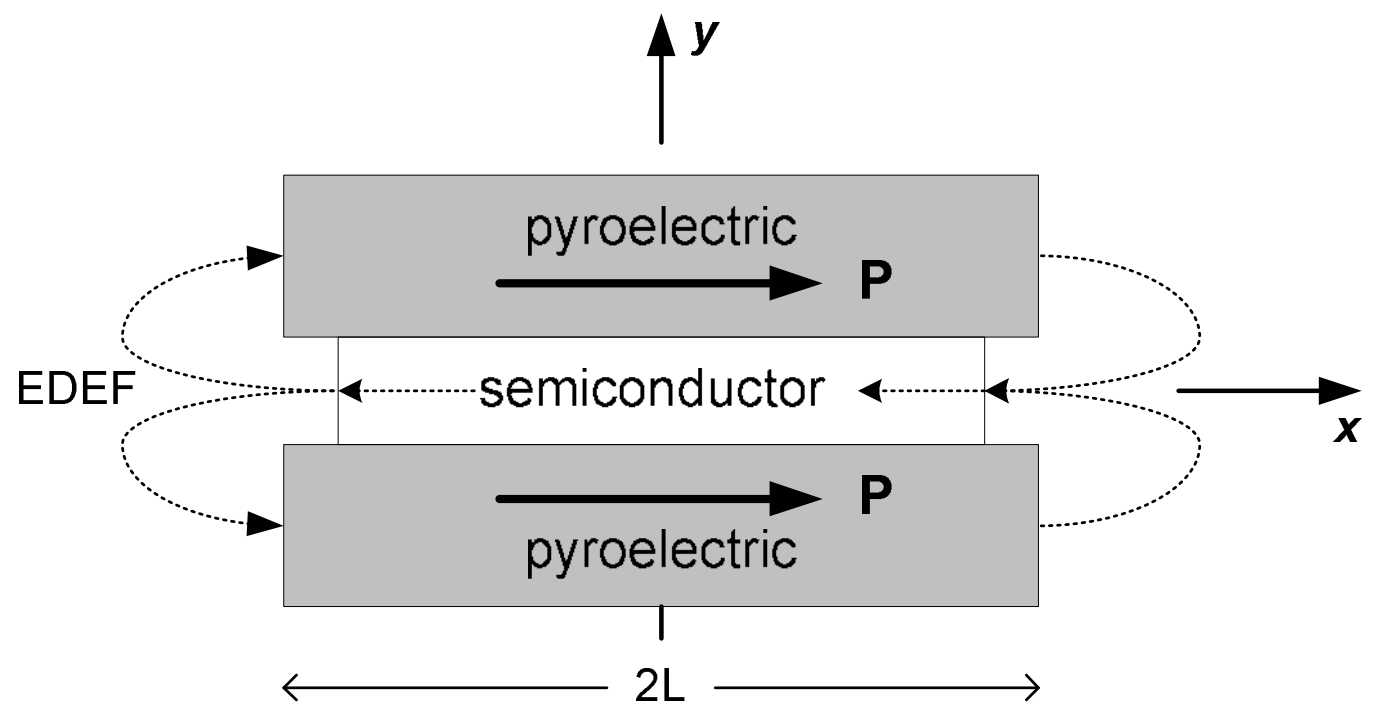

Figure 9. Schematics of the "sandwich" configuration of the PE + SC system.

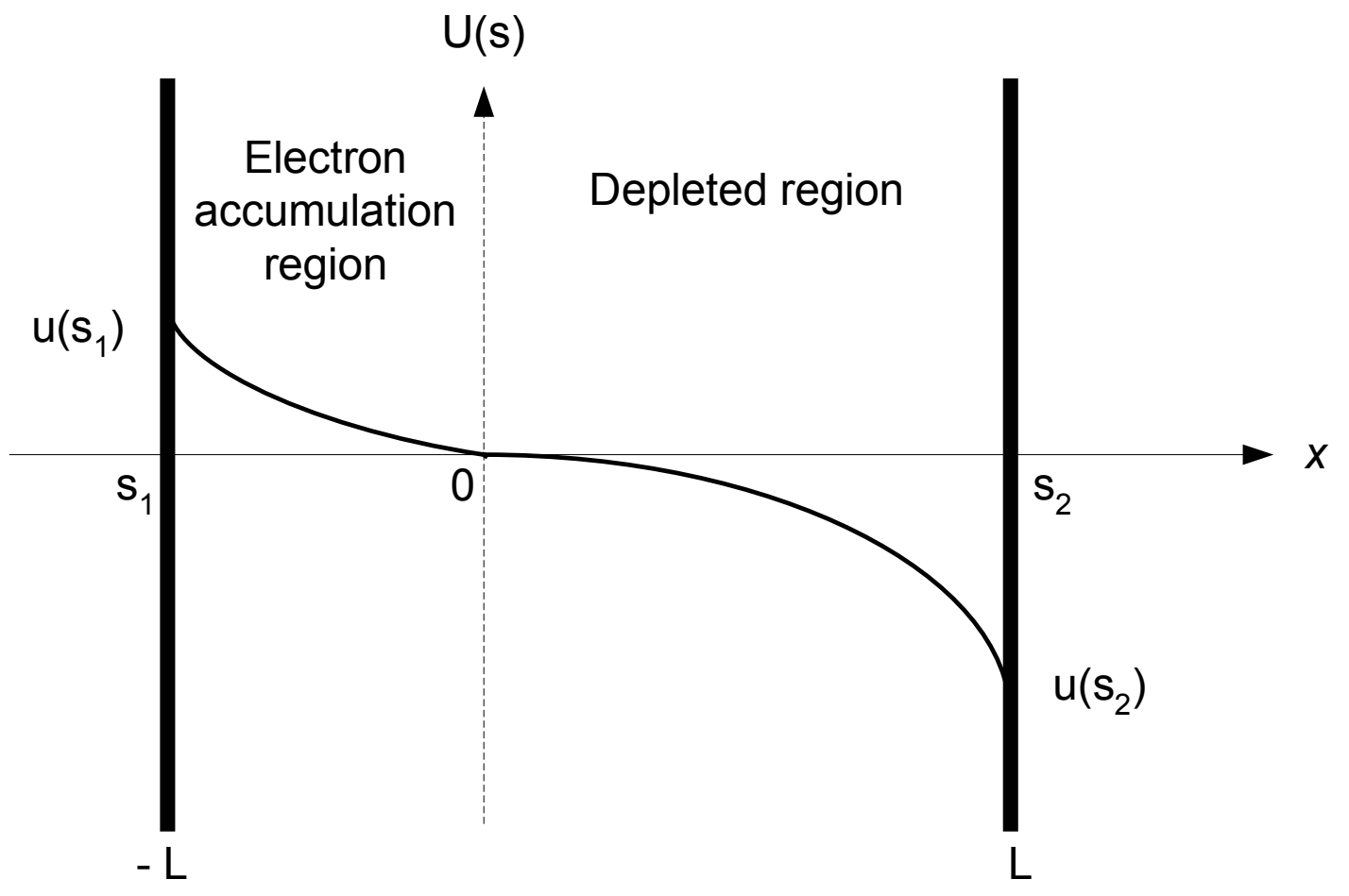

Figure 10. Graphical description of the model used in the calculation in Section 3.1

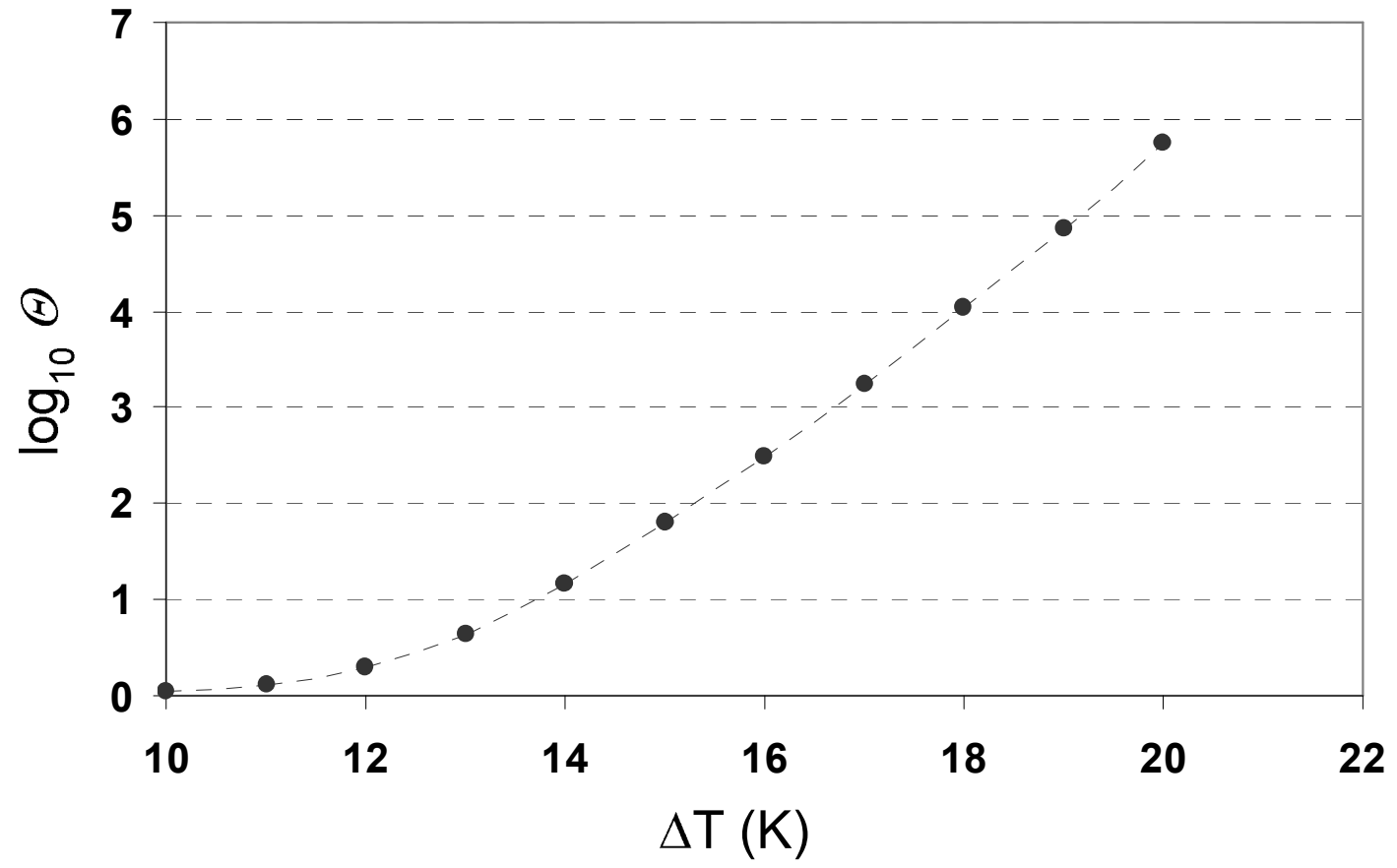

Figure 11. The dependence of the relative change of the SC resistance on the magnitude of the temperature change $\Delta T$.

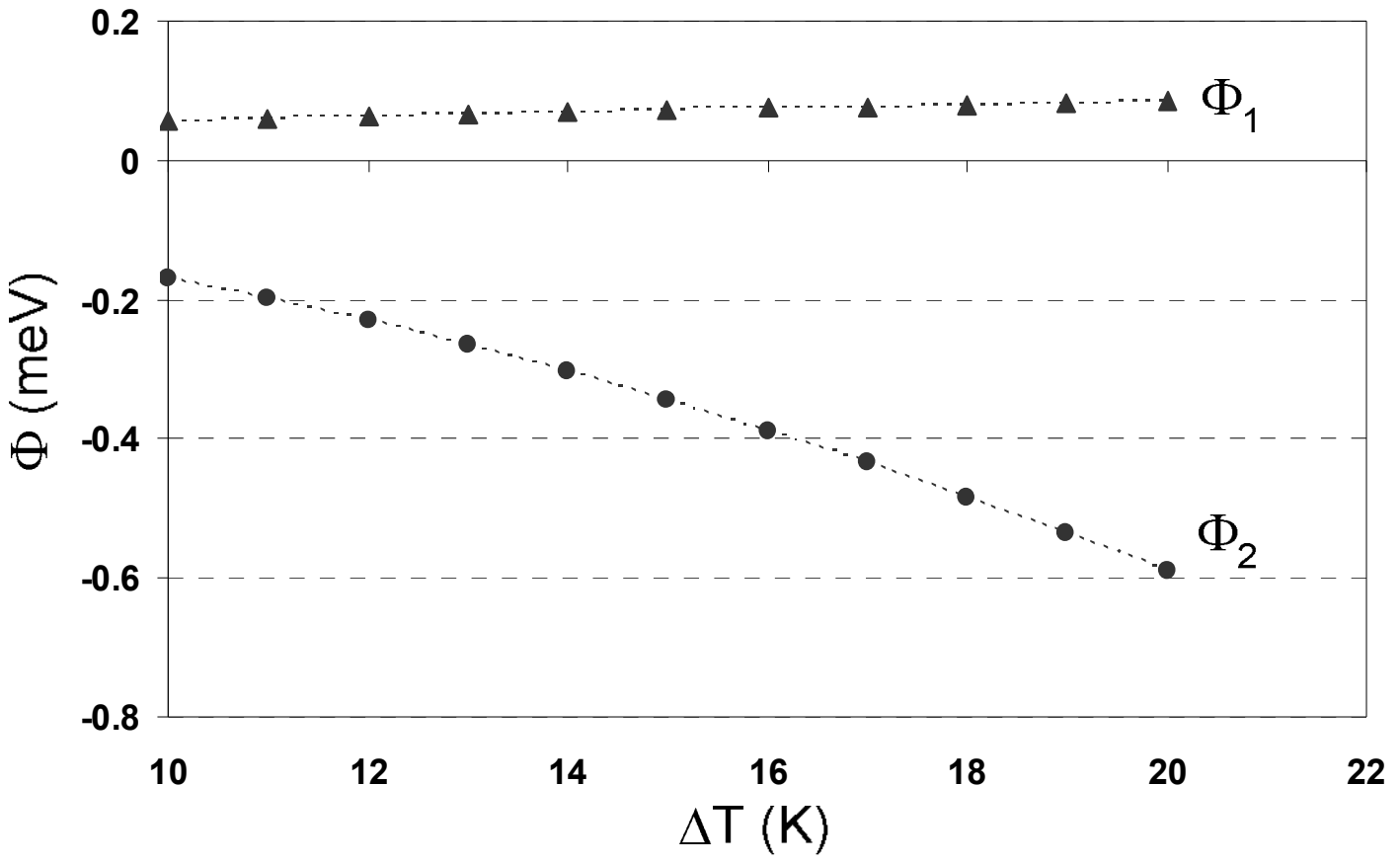

Figure 12. The dependence of the band bend $\Phi_1$ at the left, electron accumulated, "low resistive" end of SC, and of the band bend $\Phi_2$ at the right, depleted, "high resistive" end of SC, on the temperature increment.

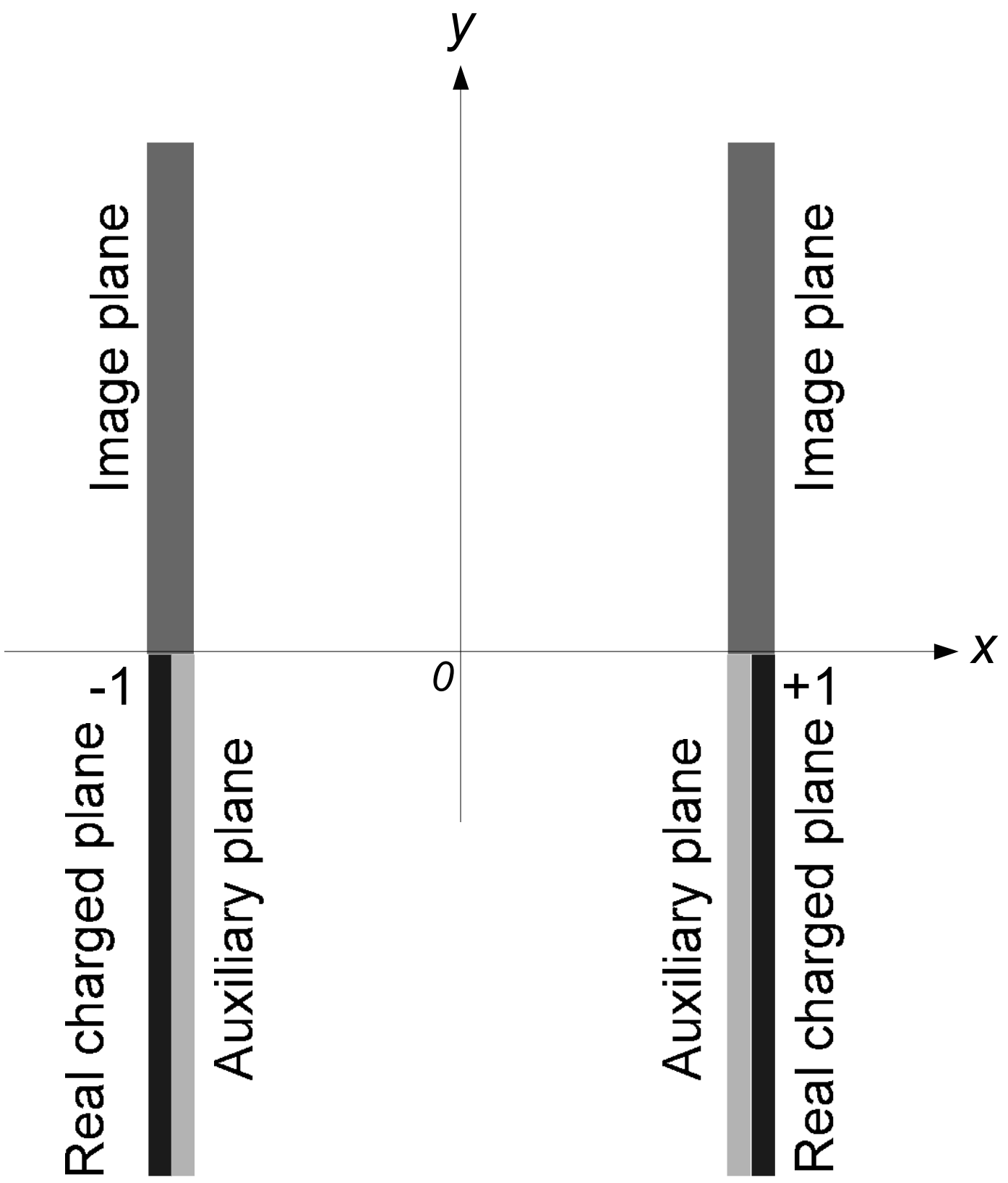

Figure 13. The description of the various real and image charged planes used in the calculations.